\documentclass{aa}  
\usepackage{graphicx}
\usepackage{indentfirst}
\usepackage{textcomp}
\usepackage{color}
\usepackage{amsmath}
\usepackage{amssymb}
\usepackage{txfonts}
\usepackage{mymacros}
\definecolor{darkgreen}{rgb}{0,0.4,0}

\begin{document} 

\authorrunning{Prabhu et al.}

   \title{Inferring magnetic helicity spectrum in spherical domains: the method and example applications}

   \author{A. P. Prabhu
          \inst{1},
          N. K. Singh\inst{2},
          M. J. K\"apyl\"a\inst{3,1,4},
          \and
          A. Lagg\inst{1,3}
          }

   \institute{Max Planck Institute for Solar System Research, Justus-von-Liebig-Weg 3, D-37077 G\"ottingen, Germany\\
              \email{prabhu@mps.mpg.de}
          \and 
          Inter-University Centre for Astronomy and Astrophysics, Post Bag 4, Ganeshkind, Pune 411007, India
          \and
         Department of Computer Science, Aalto University, PO Box 15400, FI-00076 Aalto, Finland
          \and
          NORDITA, KTH Royal Institute of Technology and Stockholm University, Hannes Alfv\'ens v\"ag 12, SE-11419, Stockholm, Sweden
             }


   \date{$ $Revision: 1.121 $ $}

    \abstract
 {Obtaining observational constraints on the role of turbulent effects for the solar dynamo is a difficult, yet crucial, task.
 Without such knowledge, the full picture of the operation mechanism of the solar dynamo cannot be formed.}
  {The magnetic helicity spectrum provides important information about the $\alpha$ effect. Here
   we demonstrate a formalism in spherical geometry to infer magnetic helicity spectra directly from 
   observations of the magnetic field, taking into account the sign change of magnetic helicity across the Sun's equator.}
   {Using an angular correlation function of the magnetic field, we develop a method to infer spectra for magnetic energy and helicity. 
   The retrieval of the latter relies on a fundamental definition of helicity in terms of linkage of magnetic flux. 
   We apply the two-scale approach, previously used in Cartesian geometry, to spherical geometry for systems where a sign reversal 
   of helicity is expected across the equator at both small and large scales.}
  {We test the method by applying it to an analytical model of a fully helical field, and to magneto-hydrodynamic simulations of a turbulent dynamo.
  The helicity spectra computed from the vector potential available in the models are in excellent agreement to the spectra computed 
  solely from the magnetic field using our method. In a next test, we use our method to obtain the helicity spectrum from a synoptic magnetic field map corresponding to a Carrington rotation. We observe clear signs of a bihelical spectrum of magnetic helicity, which is in complete accordance to the previously reported spectra in literature from the same map.}
   {Our formalism makes it possible to infer magnetic helicity in spherical geometry, without the necessity of computing the magnetic vector potential. This has the advantage of being gauge invariant.
 It has many applications in solar and stellar observations, but can also be used to analyze global magnetoconvection models of stars and compare them with observations.} 
   {}
    {}
   \keywords{
Sun: magnetic fields ---
Magnetohydrodynamics (MHD) ---
Dynamo ---
Turbulence }
   \maketitle
\section{Introduction}\label{intro}
The solenoidal nature of magnetic fields enables us to examine them in terms of the topology of 
closed curves \citep{Berger_Field}. 
Helicity integrals in general, and magnetic helicity in particular, 
(see \Eq{hel_def}) have been demonstrated to be associated with the topological properties of 
field lines \cite[]{Moffat69,Moffat78}. Magnetic helicity, which characterises the 
linkage of field lines, is a topological invariant. In ideal 
magnetohydrodynamics (MHD) magnetic helicity is conserved \citep{Woltjer}, and is 
nearly conserved in the limit of 
large conductivity \citep{Berger_84}. Thus, magnetic helicity 
imposes a crucial constraint on the evolution of magnetic fields.

In the case of the Sun, magnetic helicity is often invoked to investigate many facets of the solar magnetic field. 
To name a few examples, it is suggested as a possible proxy to quantify the 
eruptivity of solar active regions (ARs) \cite[]{Pariat2017,Thalmann2019}, 
or as a plausible predictor of the solar cycle \citep{Hawkes}. The influence of magnetic helicity on coronal emission,
to explain the observed enhancement in X-ray luminosity with increasing stellar rotation, has also been explored
\citep{Warnecke}. Arguably, magnetic helicity plays the most crucial part in dynamo theory, which is 
often called upon to explain the generation and maintenance of the solar magnetic field \citep[]{Moffat78,BS05,Bran_adv,FR}. 
Specifically the $\alpha$ effect, occurring in systems with 
stratification and rotation, is expected to be an important inductive effect 
in the solar dynamo. Under isotropic and homogeneous conditions $\alpha$ is known to be 
related to the kinetic helicity of the flow and is a measure 
of the helical nature of turbulence within the Sun's convection zone. It has been shown that the 
$\alpha$ effect generates bihelical magnetic fields, that is, magnetic helicity at small and large 
scales of opposite signs, hence resulting in no net production of magnetic 
helicity \cite[]{Seehafer96,Ji,YB03}. For the Sun, 
another sign change of magnetic helicity 
is expected across the equator, due to the Coriolis force breaking reflectional symmetry. 
Thus we have a hemispheric sign rule \cite[HSR, see][for details]{Singh}. 
The HSR implies a positive (negative) sign of magnetic helicity at large (small) scales in
the northern hemisphere and vice versa in southern hemisphere.
Observational evidence for such a sign rule will be an indirect
confirmation of the role played by the $\alpha$ effect in generating large-scale solar magnetic fields.

Magnetic helicity is defined in terms of $\AAA$, the magnetic vector potential 
and the magnetic field, $\BBB$, as
\begin{equation}
  \hhel = \int \AAA \cdot \left(\CURL \AAA \right) ~d^3 x= \int \AAA \cdot \BBB ~d^3 x.
\label{hel_def}
\end{equation}
The first hurdle in infering magnetic helicity observationally arises due to the fact that
$\AAA$ in \Eq{hel_def} is not an observed quantity and 
secondly, the information about $\BBB$ is usually not known over the 
entire volume. Given these restrictions, prior efforts  
to study the magnetic helicity of ARs used current helicity $\mathcal{H}_\mathrm{C} = \left<\JJJ \cdot 
\BBB\right>_\mathrm{V}$ as a proxy, where $\JJJ$ is the current density and the angle brackets denote volume averages. Additionally the simplifying 
assumption of a force-free magnetic field is made \cite[]{Seehafer90,Pevtsov1995,Bao99,
Zhang2010}. 
If in \Eq{hel_def} $\BBB$ has a non-zero normal component at the boundary of the 
integration volume, then the volume integral cannot be determined in a gauge invariant manner.
This is true for most astrophysical systems including the Sun. For such situations 
the concept of relative magnetic helicity can prove useful \cite[]{Berger_Field,Finn}:  
in this approach, helicity is defined with respect to a reference field, usually a current-free 
one.
Relative magnetic helicity is often used to analyse solar observations, 
as is demonstrated in a recent 
study by \cite{Thalmann2019}. Using nonlinear force-free field extrapolations as input, these 
approaches involve an explicit computation of $\AAA$ making suitable gauge choices. 
For a review of such and some other related methods see \cite{Valori16}.

Using a toroidal-poloidal decomposition \citep{Chandra} of vector 
fields in spherical geometry, \cite{Pipin2019} compute the magnetic helicity from solar synoptic
maps by reconstructing $\AAA$, again making a particular gauge choice. 
In their analysis they separate the large and small scales relying on azimuthal averages,
wherein the small-scale magnetic helicity density includes contributions from all scales 
except the axisymmetric mean field, including the large-scale non-axisymmetric contributions.
\cite{Lund} also resort to such a decomposition, applying it on stellar data retrieved 
via Zeeman-Doppler imaging \cite[ZDI,][]{Semel}, although they report large-scale average 
magnetic helicity density on stellar surfaces. 
However, in these methods, the total value of magnetic helicity density does not provide us the distribution 
of helicity, specifically its sign, over scales.

To study the dynamo problem, for which the segregation of helicity over spatial scales is expected, methods 
focused on inferring spectral
distribution of magnetic helicity from observations are better suited.
\cite{Zhang_2014,Zhang_2016} 
inferred such a spectrum of magnetic helicity from local patches, invoking the two-point correlation
tensor for magnetic field in Cartesian geometry. \cite{Brantwo} extended this approach to 
inhomogeneous systems by using the \textit{two-scale analysis} of \cite{Roberts_soward}, applicable 
to cases where magnetic helicity varies slowly, on a 
scale somewhat smaller than the scale of the system. It is important to note that these methods
avoid any issues with a gauge choice by transforming into Fourier space with the implicit assumption
of periodicity. Recently \cite{Prior} demonstrated a wavelet based multi-resolution analysis to infer
magnetic helicity which is applicable for inhomogeneous systems with non-periodic boundaries. 
They also avoid computing the vector potential and thus avoid 
making a gauge choice by relying on a topologically meaningful definition of helicity 
in terms of $\BBB$. The investigation of the helicity distribution on global scales, i.e. over the whole surface of
the Sun or other stars, requires the use of a spherical geometry.
To investigate the helicity distribution over scales on the surface of the Sun or other stars,
the observations are available in spherical geometry.
In this context \cite{Brantwo} regarded their 
Cartesian approach as preliminary and highlighted the need to do an analogous analysis in spherical harmonics.
\cite{Prior} also assume a Cartesian volume for their multi-resolution wavelet decomposition.

In this paper we rely on a similar topological definition of helicity in terms of linking of
$\BBB$, and adapt it to spherical geometry by using spherical harmonics.
Here we extend the Cartesian approach of \cite{Brantwo} and do a treatment of scales based 
on spherical harmonic degree in contrast to the approach of \cite{Pipin2019} which is based on azimuthal averaging. Our method is
suited for solar and stellar observations, where photospheric magnetic field observations are 
available in spherical geometry via spectropolarimetric inversions or ZDI. Particularly for the Sun, it also enables
access to higher cadence full disk magnetogram data. It is also readily applicable to solar/stellar dynamo models,
where the induction equation is typically expressed in $\BBB$ and numerically integrated, 
to compute magnetic helicity from such models and compare them with observations, 
where $\AAA$ is not available.

This paper is organized as follows:
In Sect.~\ref{theory} we describe our present 
method for spherical geometry in the context of the existing Cartesian framework. 
In Sect.~\ref{test}, we test this formalism against an analytical expression of a magnetic field on a 
spherical shell. Moreover, we also apply it to a three-dimensional turbulent dynamo 
simulation in spherical geometry. Finally we also apply a simple modification of our approach 
to synoptic maps of the Sun and compare it with previously obtained results. 
We discuss the scope of applicability of our formalism to observations in Sect.~\ref{discuss}

\section{The method and formalism} \label{theory}

\subsection{Definitions}\label{theory_def}
We begin by defining the spectra for magnetic energy and helicity on a spherical shell.
In such situations it is convenient to use spherical harmonics,
$\ylm(\theta,\phi)$, where $\ell$  
represents the degree and $m$ the azimuthal order. Thus we can define the spectra of magnetic energy 
as,
\begin{equation}
    2E_\mathrm{M} (\ell) = \sum_{m=-\ell}^{\ell} b_i^{\ell m}
	b_i^{\ell m*},\label{en_spec}
\end{equation}
where $b_i^{\ell m} = \int B_i(\theta,\phi)~ \ylm(\theta,\phi) ~d\Omega$ is the expansion coefficient of the $i^\mathrm{th}$ component of $\BBB$, 
$i = (r,\theta,\phi)$,
and the star symbol ($*$) denotes the complex conjugate of the coefficient. 
On such a spherical shell 
we have $\int d\Omega = 4\pi$ as the surface area of a unit sphere. Analogously, following \Eq{hel_def},
 we can define a spectrum of magnetic helicity as
\begin{equation}
    H_\mathrm{M} (\ell) = \sum_{m=-\ell}^{\ell}
	\frac{1}{2}\left[a_i^{\ell m*~}b_i^{\ell m} +
	a_i^{\ell m}~b_i^{\ell m*}\right].\label{hel_spec}
\end{equation}
These spectra can be directly related to the values of magnetic energy and helicity using 
Plancherel's theorem:
\begin{align}
   \mathcal{E}_\mathrm{M} &\equiv \frac{1}{4\pi}\int\frac{1}{2}
	\BBB^2~d\Omega = \sum_{\ell=0}^{\infty}E_\mathrm{M} (\ell), {\ \ \rm and}\label{plan_e}\\
    \mathcal{H}_\mathrm{M} &\equiv \frac{1}{4\pi}\int \AAA \cdot
	\BBB~d\Omega = \sum_{\ell=0}^{\infty}H_\mathrm{M} (\ell)
	\label{plan_h}.
\end{align}

As has been demonstrated in \cite{Moffat69,Berger_Field}, for a given field $\BBB$,
we can define the magnetic helicity in terms of the linkage of its flux, thus using the Gauss linking formula :
\begin{equation}
      \mathcal{H}_\mathrm{M} = \frac{1}{4\pi} \int \int \BBB(\xxx) \cdot\left[\BBB(\yyy) \times \frac{\xxx-\yyy}{|\xxx-\yyy|^3} \right]~ d^3x~ d^3y. \label{gauss_link}
\end{equation}
This definition is equivalent to that of \Eq{hel_def} if one adopts the Coulomb gauge, $\DIV \AAA =0$,
and uses the Biot-Savart law \citep[]{Moffat_Ricca,Sub_helden}. \cite{Sub_helden} 
demonstrate how this definition allows for a gauge-invariant, physically meaningful description of 
magnetic helicity density.

\subsection{Cartesian geometry}\label{theory_cart}
\noindent
\emph{(a) Homogeneous case:}
In this section we first reiterate the approach delineated in
\cite{Moffat78} that relates the correlation function of the magnetic
field to its energy and helicity in homogeneous conditions.
The magnetic energy spectrum, $E_{\rm M}(k)$, is obtained from
$2E_{\rm M}(k)=\int \delta_{ij}\hat{M}_{ij}(\kkk)\,k\,\dd\Omega$,
where $\hat{M}_{ij}(\kkk)$ is the 2D Fourier transform of the two-point
correlation tensor of the total magnetic field, i.e.,
$M_{ij}(\xxi)=\brac{B_i(\xxx)B_j(\xxx+\xxi)}$, which depends on the
separation $\xxi$ between the two points.
This tensor is assumed to be independent of the position vector $\xxx$ under homogeneous
conditions; the brackets denote an ensemble average and $\kkk$ is
the conjugate variable to $\xxx$ spanning the 2D Cartesian surface.
The scaled magnetic helicity spectrum, $kH_{\rm M}(k)$, with
same dimensions as that of $E_{\rm M}(k)$, is defined as
$kH_{\rm M}(k)=\int\ii\hat{k}_i\epsilon_{ijk}\hat{M}_{jk}(\kkk)\,k\,
\dd\Omega$, with $\hat{k}_i=k_i/|\kkk|$ being the unit vector of $\kkk$.
\\

\noindent
\emph{(b) Inhomogeneous case:}
This formalism was extended to inhomogeneous 
conditions in \cite{Brantwo}. Under such conditions, 
for a given magnetic field $\BBB(\xxx,t)$, 
with $\xxx$ being the position vector at a time $t$,  
its two-point correlation tensor 
in Fourier space takes the form of
\begin{equation}
\tilde{M}_{ij}(\KKK,\kkk) = \left<\hat{B}_i(\kkk+\frac{1}{2}\KKK) ~\hat{B}_j^{*}(\kkk-\frac{1}{2}\KKK)\right>.\label{cart_2k} 
\end{equation}
Here $\KKK$ is the conjugate variable to
$\XXX = (\xxx'+\xxx'')/2$, the slowly varying coordinate, 
and $\kkk$ the conjugate to $\xxx = \xxx'-\xxx''$, 
the distance between the two points around $\XXX$. For details see \cite{Roberts_soward,Brantwo}.
For brevity we henceforth omit specifying explicitly the time dependence.
Then the spectrum for magnetic energy and helicity are given by
\begin{align}
   2\tilde{E}_\mathrm{M} (\KKK, k) &= \int \delta_{ij}~\tilde{M}_{ij}(\KKK,\kkk)~k ~d\Omega, \label{en_spec_c}\\
   k\tilde{H}_\mathrm{M} (\KKK,k) &= \int i\hat{k}_i~\epsilon_{ijk}~\tilde{M}_{jk}(\KKK,\kkk)~k ~d\Omega \label{hel_spec_c},
\end{align}
where $\int d\Omega = 2\pi$. Therefore, the trace of $\tilde{M}_{ij}$ gives the magnetic energy and its 
skew-symmetric part the magnetic helicity. 
Equation~(\ref{hel_spec_c}) is related to the definition 
of helicity in \Eq{gauss_link}. The spectrum is thus computed by the product of magnetic fields that 
are shifted by a wavenumber corresponding to the large-scale modulation of the slowly varying 
coordinate. We will draw upon this idea in Sect.~\ref{test_dyn}.

\subsection{Spherical geometry}\label{theory_sph}

\noindent
\emph{(a) Homogeneous case:}
For a quantity distributed on a sphere, the use of an angular correlation function to extract its power
spectrum has been demonstrated by \cite{Peebles} in the context of cosmology.
Here we apply these principles to an angular correlation tensor of the total magnetic field vector on 
the observed surface of a star.
For the magnetic field vector at two positions $\OO_1$ and $\OO_2$ on a unit sphere, the correlation 
function is of the form (see Appendix~\ref{app1} for details),
\begin{align}
    M_{ij}(\chi)&= \int \frac{1}{8\pi^2} B_i(\OO_1) B_j(\OO_2) \delta(\cos\chi_{12}-\cos\chi) ~d\Omega_1 d\Omega_2 \label{corr_fn},
\end{align}
with $\OO = (\theta,\phi)$, and $\theta$ and $\phi$ are the colatitude and azimuth respectively and 
$\chi_{12}$ is the angle between the two directions $\OO_1$ and $\OO_2$. Here 
we assume homogeneity, where $M_{ij}$ is assumed to depend only on the separation $\chi$. The delta 
function ensures the product is evaluated at the angular separation and the denominator is needed for
normalisation \citep{Peebles}. Now, analogous to the Cartesian case, the spectrum of magnetic energy
is given by the trace of the correlation tensor as (see Appendix~\ref{app2})
\begin{equation}
    2E_\mathrm{M}(\ell) = 2\pi (2\ell+1) \int \delta_{ij} M_{ij} (\chi) ~ 
	P_{\ell}(\cos\chi) ~\dd\cos\chi,
    \label{corr_fn_e}
\end{equation}
and the scaled spectrum of
magnetic helicity is given by its skew-symmetric part (see Appendix~\ref{app3}),
\begin{equation}
	k H_\mathrm{M}(\ell) = 2\pi \int \epsilon_{ijk}
	\nabla'_i (P_\ell(\cos\chi) M_{jk} (\chi) ~\dd\cos\chi
   \label{corr_fn_h}.
\end{equation}
Here, $\nabla'\equiv \partial/\partial\theta_2+(1/\sin\theta_2)\,\partial/\partial\phi_2$
acts on location $\OO_2(\theta_2,\phi_2)$, and $\chi$ is the angular
separation between $\OO_1$ and $\OO_2$; see Appendix~\ref{app3} for
a derivation and note that $k R \approx \ell+1/2$ by the Jeans relation
where $R$, taken here as unity, is the radius of the sphere.
We use the general expression given in \Eq{final_exp},
which directly determines the linkages of magnetic field lines to
yield the magnetic helicity, and is also applicable to the inhomogeneous
case discussed below. \Equ{corr_fn_h}, as written above, may be obtained by
first taking the ensemble average of \Eq{final_exp} and then using the
homogeneous form of $M_{ij}(\chi)$ given in \Eq{corr_fn}.
\\

\noindent
\emph{(b) Inhomogeneous case:}
We focus here on slow latitudinal variation of magnetic helicity which
has opposite signs in the two hemispheres; see \cite{Singh} for the
expected HSR of the Sun. As noted in Appendix~\ref{app3}, the two-point
correlation function $M_{ij}(\OO,\chi)$ depends on both, the position on the sphere
as well as the separation between the two points; $\OO$ here represents
a mean location that lies between the two points.
In exact analogy to the Cartesian case discussed earlier in
\Sec{theory_cart}, we find that the inhomogeneous case corresponds
simply to a shift in the spherical degree $\ell$ by, say, $\eell$,
while determining the spherical transform of $M_{ij}$; see
\Eqs{cart_2k}{hel_spec_c} for the analogy, and \cite{Bran_global}
where this generalization in the spherical domain was discussed.
The scheme to determine the helicity in this case may be briefly
sketched here as: $H_{\rm M}(\eell,\ell)\propto \brac{b_i^{\ell m} b_j^{\ast\, \ell+\eell,m}}$,
where $b_i^{\ell m}$ and $b_j^{\ast\, \ell+\eell,m}$ correspond to the
two components of the magnetic field.
We have chosen $\eell=1$ in the present study where it corresponds to the
large scale modulation of the magnetic helicity; as
$P_1(\cos\theta)=\cos\theta$, it naturally involves a sign change
across the equator where $\theta=\pi/2$.
To show the spectrum of magnetic helicity in the sections below, we
adopt a sign convention that corresponds to the northern hemisphere,
where the large-scale fields have positive helicity.

\section{Testing the method}\label{test}
To test whether our formalism allows us to infer the spatial distribution magnetic helicity over a 
sphere relying only on $\BBB$, we subject it to cases where we know both $\AAA$ and $\BBB$. In 
Sect.~\ref{test_CK}, we use an analytical expression of a fully helical magnetic field, and in Sect.~\ref{test_dyn}
a simulated magnetic field generated by a turbulent dynamo.
Finally, we apply it to synoptic vector magnetograms of the Sun in Sect.~\ref{test_obs}.
  
\begin{figure}
    \centering
    \includegraphics[width=\columnwidth]{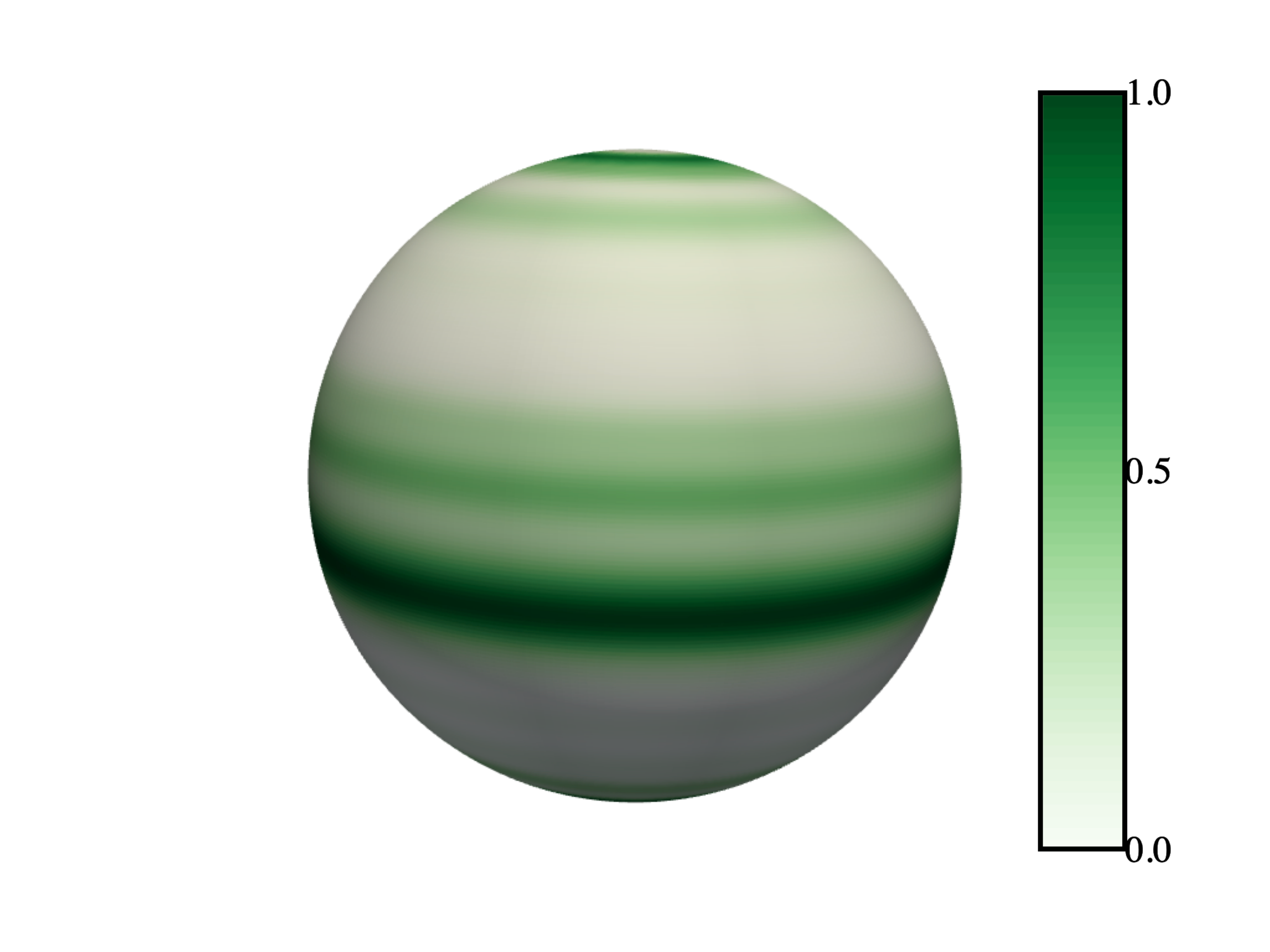}
    \caption{
    Distribution of magnetic helicity density obtained from $\AAA$, defined in \Eq{ck_pot}, and 
    $\BBB=\alpha\AAA$. The color represents its magnitude which has been normalised by the maximum value.
    This figure
    was made using PyVista \citep{sullivan2019pyvista}. The grey tinge is simply a shading effect.}
    \label{fhel_dis}
\end{figure}

\subsection{Fully helical field}\label{test_CK}
\begin{figure}
    \centering
    \includegraphics[width=\columnwidth]{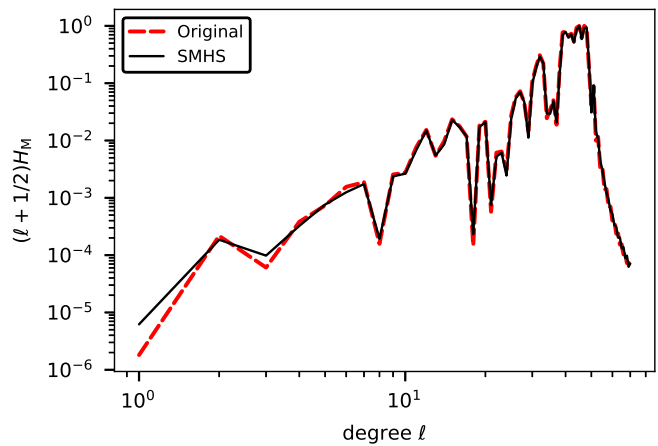}
    \caption{Comparison of the original spectrum (red dashed) obtained using 
    \Eq{hel_spec} for the helical magnetic field defined in Sect.~(\ref{test_CK}) versus the spectrum obtained with 
    \Eq{corr_fn_h} with only the knowledge of magnetic field. Both spectra are normalised to their maximum value.}
    \label{fhel_comp}
\end{figure}

\cite{CK_1957} derived general solutions of the equation $\CURL{\BBB}=\alpha\BBB$, in the context of
force-free magnetic fields for spherical geometry
 commonly known as Chandrasekhar-Kendall functions. 
Here we use these functions, $\TTT = \CURL{(\phi \hat{r})}$ and 
$\SSS = \CURL{\TTT}/\alpha$, to define the vector potential on a unit sphere as 
$\AAA = \TTT+\SSS$. We expand $\phi$ in spherical harmonics as
$\phi = \sum_{\ell m} c^{\ell m}\ylm{(\OO)}$, 
assume axisymmetry and ignore the radial dependence of the
coefficients $c^{\ell m}$ while  choosing their values in a random fashion. 
Thus we have the vector potential as,
\begin{align}
    \AAA (\OO)&= \frac{-1}{\alpha}L^2\phi~\ru  - \frac{\pd \phi}{\pd \theta}~\rphi,\nonumber\\
     &= \frac{1}{\alpha}\sum_{\ell m}\ell(\ell+1)c^{\ell m}\ylm(\theta,\phi)~\ru  - \sum_{\ell m} c^{\ell m} \frac{\pd}{\pd \theta} \ylm(\theta,\phi)~\rphi. \label{ck_pot}
\end{align}
Here, $L^2$ is the angular part of the Laplace operator in spherical geometry such that 
$L^2 \ylm = -\ell(\ell+1)\ylm$ and $\alpha = \sqrt{\ell(\ell+1)}$.
From this the magnetic field follows as 
$\BBB = \CURL{\AAA} = \alpha (\TTT+\SSS)$. To evaluate this numerically
we use the Python interface of SHTOOLS \citep{SHtools}. Figure~\ref{fhel_dis} shows the distribution
of magnetic helicity obtained from such a configuration. We apply \Eq{corr_fn_h} only to the magnetic field and 
compare the magnetic helicity spectra thus obtained with the one that we have a
ready access. Figure~\ref{fhel_comp} shows that
it is possible to retrieve the spherical magnetic helicity spectrum (SMHS) over 
all scales using the angular correlation function to a very high degree.

\subsection{Magnetic helicity from simulations of a turbulent dynamo}\label{test_dyn}

Another suitable test case is a simulated magnetic field from a model, where dynamo action occurs in a 
turbulent fluid in a spherical domain ($r,\theta,\phi$). We employ 3D hydromagnetic simulations of an isothermal gas where 
turbulence is driven by forcing the momentum equation with a helical forcing function
using the Pencil Code \citep{Pencil}. Along with the continuity and momentum equation, 
the Pencil Code solves the induction equation for $\AAA$, 
\begin{equation}
\dert{\AAA} = \UUU \times \BBB - \eta \mu_0 \JJJ - \nabla \psi,
\end{equation}
where $\UUU$ is the velocity field, $\eta$ is the magnetic resistivity,
$\psi$ is the electrostatic potential and $\mu_0$ is the magnetic permeability.
This formulation ensures the solenoidality of $\BBB$, thus the vector potential is readily available with
the Pencil Code. Here we use the resistive gauge, $\psi=\eta\nabla\cdot\AAA$, for our simulations.

The simulation domain spans $0.7R_0\leq r\leq R_0$ in radius, to mimic the convection zones of solar-like
stars. Its extent in colatitude is $2\pi/5\leq\theta\leq3\pi/5$ and $\pi/2$ 
in the azimuthal direction, hence our simulation domain is wedge shaped.
We use periodic boundary conditions in the azimuthal direction. For velocity,
stress-free and impenetrable boundary conditions are used for both boundaries
in the radial and latitudinal direction. For the magnetic vector potential, 
at both latitudinal boundaries and at the bottom, perfect conductor boundary conditions are used, 
whereas at the top boundary a radial field condition is used.
These conditions allow for magnetic helicity fluxes out of the system.

A more detailed description of the model, including the forcing function,
can be found in \cite{Warn_model}, with a key difference being that
we only have a single layer model in radius, implying that the forcing is applied at all radial locations.
We limit the extent in $\theta$ and $\phi$ for computational reasons, 
and resolve our model with a grid of $128\times256\times256$ points in radius, 
latitude and longitude, respectively. We apply the forcing at a length scale ten times smaller than the 
radial extent with maximally helical forcing. We produce two different models: in Run A we keep the sign 
of forcing the same for both hemispheres, corresponding to a negative kinetic helicity
($\UUU \cdot \www~\text{with}~\www = \CURL{\UUU}$) over the entire domain, while
in Run B we vary its sign with $\cos(\theta)$, mimicking the hemispheric sign rule expected for the Sun.

\begin{figure}
    \centering
    \includegraphics[width=\columnwidth]{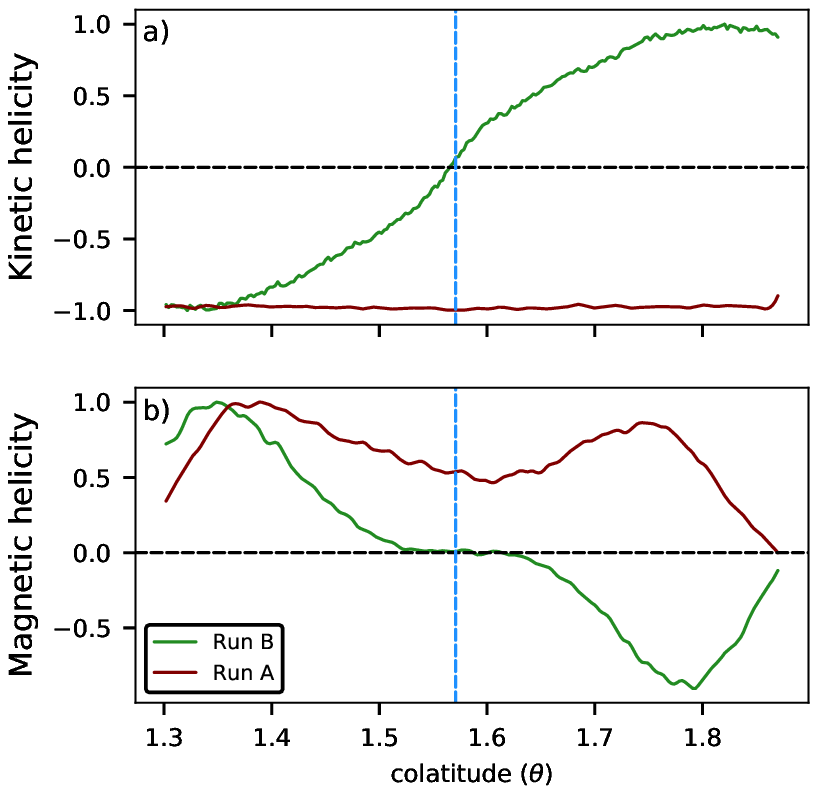}
    \caption{ Profiles of $\left<\uuu\cdot \www \right>_{\phi t}$ (panel a) and $\left<\AAA \cdot \BBB \right>_{\phi t}$ (panel b)
    as a function of $\theta$. The profiles have been normalised to their maximum value. The blue dashed line indicates the equator.}
    \label{dyn_profs}
\end{figure}

The magnetic Reynolds number, $\Rm$, a non-dimensional parameter quantifying the effects magnetic 
advection to diffusion, is 9.5 for Run A and 19.2 for Run B. The magnetic Prandtl number, $\Pm=\nu/\eta$, for
both runs is unity. Here $\nu$ is the kinematic viscosity.
Figure~\ref{dyn_profs}a shows the $\phi$- and time-averaged profiles of kinetic helicity at 
$r_1 \approx0.81R_0$. For Run A it is negative over the entire domain, and for Run B, 
it changes sign across the equator from negative to positive from northern to southern hemisphere. 
For such helically forced dynamos, the magnetic helicity is expected to be dominated by large-scale 
fields and its sign is expected to be opposite to that of kinetic helicity \citep{Bran_inv}. 
This behaviour is reflected in our simulations as seen in Fig.~\ref{dyn_profs}b  for both Runs A and B.

\begin{figure*}
    \centering
    \includegraphics[width=2\columnwidth]{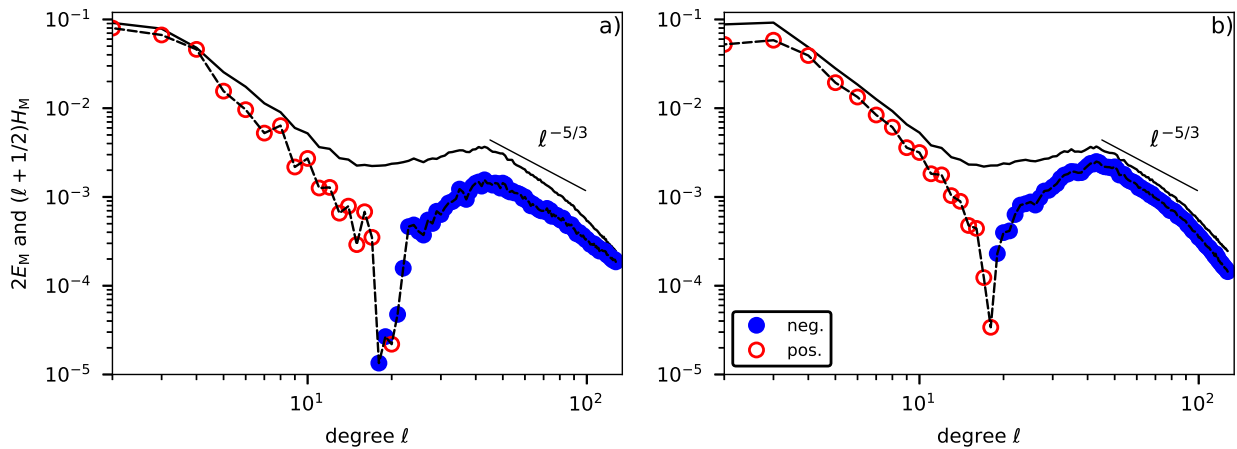}
    \caption{Magnetic energy (solid) and scaled helicity spectra (dashed) obtained at a depth of $r_1$ for Run A. In panel
    a) the energy and helicity spectra are obtained using the magnetic field and vector potential directly from the simulation data,
    whereas in panel b), we use the angular correlation function given by \Eq{corr_fn} to evaluate the energy spectra and
    the SMHS using only the magnetic field data.}
    \label{corrfn_w_sc}
\end{figure*}

At first we focus on Run A, which is homogeneously forced and thus our formalism in 
Sect.~\ref{theory_sph}(a), assuming homogeneity, is applicable to it. 
Figure~\ref{corrfn_w_sc}a shows the magnetic energy and scaled helicity spectra obtained
directly from the simulation using the magnetic field and vector potential as input.
We use \Eqs{en_spec}{hel_spec} to compute these spectra. In contrast, the energy and scaled helicity spectra shown
in Fig.~\ref{corrfn_w_sc}b are obtained using only the magnetic field with \Eqs{corr_fn_e}{corr_fn_h}.
This shows that SMHS (Fig.~\ref{corrfn_w_sc}b) computed using the angular correlation function of magnetic field 
recovers a bihelical spectrum with positive (negative) sign at large (small) scales, which is in very good agreement
with the actual helicity spectrum of the simulation (Fig.~\ref{corrfn_w_sc}a). All spectra are computed at a depth of $r_1$ at one instance in time,
and then averaged over time after the large-scale field has saturated.
We note that the realisability condition of $|H_\mathrm{M}|\leq 2r_1E_\mathrm{M}/(\ell+1)$ \cite[]{Moffat78,Kahniashvili} is met. 
The energy spectrum shows an approximate $\ell^{-5/3}$ behaviour below
the injection scale ($\ell\approx 42$). 
Following \Eq{plan_h}, the ratio of magnetic helicity 
computed in real space to that of spectral space is $\approx1.06$ which points to a reasonable 
agreement. 
We note that the strong "dip" at $\ell \approx$ 20, recovered by both methods, could indicate
that in these models helicity fluxes out of the domain occur at these scales. 
This is allowed by the boundary conditions, but a thorough analysis of this phenomenon is out of the scope of the present study.

As mentioned above, this simulation was
done with a choice of resistive gauge. Therefore, to test the 
robustness of our results against the gauge choice made, we performed an 
additional run identical to Run A but with a Weyl gauge, $\psi=0$.
Using the Weyl gauge, the SMHS (not shown here) retrieved using the angular correlation function of magnetic field is bihelical. 
And as above, the ratio of helicity computed in real space to spectral space is $\approx1.03$, thus highlighting the
gauge-invariance of our approach.

\begin{figure}
    \centering
    \includegraphics[width=\columnwidth]{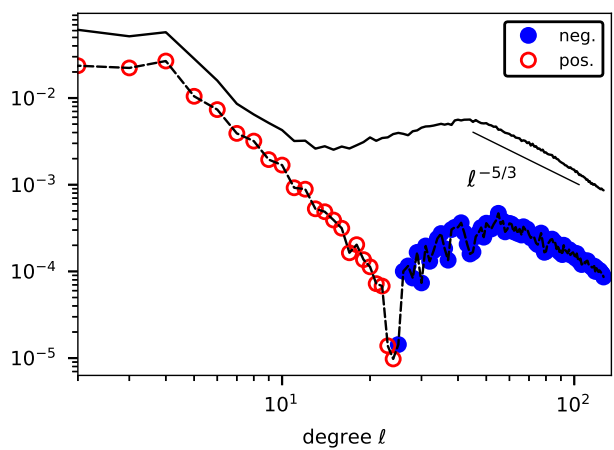}
    \caption{Magnetic energy (solid) and scaled helicity spectra (dashed) for Run B. Both spectra are obtained using
    the angular correlation function shown in \Eq{corr_fn}. The energy spectrum is obtained
    using \Eq{corr_fn_e} and the SMHS with the formalism described in Sect.~\ref{theory_sph}(b) }
    \label{corrfn_w_sc_prof}
\end{figure}

For Run B, the inhomogeneous forcing results in magnetic helicity 
slowly changing sign as a function of colatitude at both large and small scales, where Fig.~\ref{dyn_profs}a reflects the 
opposite signs of helicity at large-scales in the two hemispheres. Such behaviour is expected in the 
Sun and other stars. For magnetic field extracted at $r_1$ from Run B, we applied the formalism presented in
Sect.~\ref{theory_sph}(b) and the spectra thus obtained are shown in Fig.~\ref{corrfn_w_sc_prof}. It indeed allows us to extract the bihelical 
spectrum of magnetic helicity with signs corresponding to that of the north hemisphere, even though
the helicity changes sign as a function of latitude. As an additional confirmation, following 
\Eq{plan_h} we computed the ratio of helicity in real space to spectral once again, but computing the 
magnetic helicity in real space  over the north hemisphere only. We retrieve a value of $\approx 1.06$ from the spectrum in 
Fig.~\ref{corrfn_w_sc_prof} which is again proving the accuracy of our method.

\subsection{Solar observations}\label{test_obs}
As a final test of the applicability of the formalism put forth in Sect.~\ref{theory_sph}(b) 
to inhomogeneous systems, we apply it on synoptic vector magnetograms of the Sun, 
where a change in sign of helicity depending on the hemisphere is expected. \cite{Singh} 
studied 74 synoptic Carrington rotations (CRs) maps, using the Cartesian two-scale formalism, 
based on Vector Spectromagnetograph (VSM) data of the SOLIS project \cite[]{Keller,Bala}.
They recovered  a bihelical spectrum in a majority of cases studied.
We choose here the CR 2156 of the same dataset, close the the maximum of the solar cycle 24. 
This CR was reported to show a clear bihelical spectrum, with the signs at large and small scales following the
HSR, like the majority of other CRs during cycle 24. However, it was found to be peculiar
in the sense that it shows higher power and a positive sign of helicity at intermediate scales,
in comparison to the other CRs where a negative sign of helicity was prominent at these scales.
\cite{Singh} interpreted this as a sign of the dominance of the large-scale magnetic field
due to its "rejuvenation" close to the solar maximum.

Even though higher resolution data is available, we chose here to continue using SOLIS data
to enable a better comparison. This way we avoid instrumental and data reduction related differences such 
as varying instrumental resolution and disambiguation methods needed 
for resolving the $180^\circ$ ambiguity of the transverse (perpendicular to line-of-sight) component 
of the magnetic field. Further details of the SOLIS synoptic map used here can be found in \cite{Singh}.

\begin{figure*}
    \centering
    \includegraphics[width=2\columnwidth]{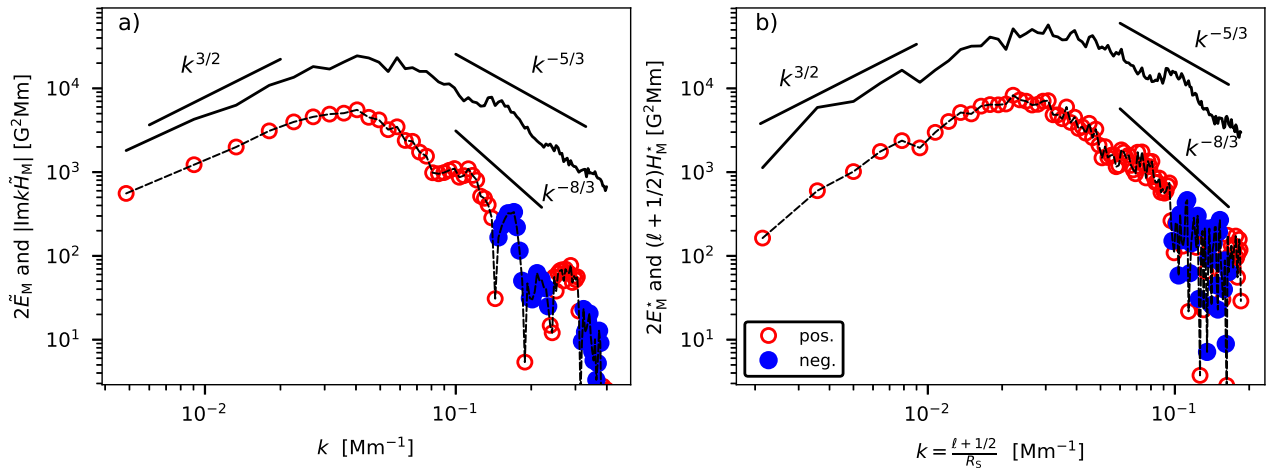}
    \caption{Magnetic energy (solid) and scaled helicity (dashed) spectra for CR 2156 obtained using SOLIS/VSM data. 
    In panel a), we have reproduced Fig.~$7$ of \cite{Singh} using . In panel b), the magnetic and SHMS obtained using the 
    angular correlationg function in spherical geometry, see Sect.~\ref{theory_sph}. Both
    plots are shown as a function of wavenumber using the Jeans relation $kR_\mathrm{S} \approx \ell + 1/2$, where $R_\mathrm{S}$ is the
    solar radius taken to be approximately $700$\,Mm. To enable a 
    better comparison between the figures we have plotted $E_\mathrm{M}^\star(\ell) = R_\mathrm{S}E_\mathrm{M}(\ell)$ and 
    $H_\mathrm{M}^\star(\ell) =R_\mathrm{S}H_\mathrm{M}(\ell)$ in panel b).}
    \label{corrfn_solis_lr_2156}
\end{figure*}

In Fig.~\ref{corrfn_solis_lr_2156}a, we reproduce Fig.~$7$ of \cite{Singh} for CR 2156 using the Cartesian two-scale formalism, 
that is \Eqs{en_spec_c}{hel_spec_c}. In panel b) of the same figure, we show the magnetic energy and SMHS of the same
CR, computed using \Eq{corr_fn_e} and the steps described in Sect.~\ref{theory_sph}(b) respectively.
The spectra shown in these two panels are in very good agreement and  reveal the bihelical nature of the solar magnetic field.
The magnetic helicity for this CR peaks at $209-300$\,Mm which is an even larger scale than the $160$\,Mm reported by \cite{Singh}. This 
lends support to their rejuvenation argument. We note, however, that SMHS recovers somewhat less 
power and energy at the largest scales than the Cartesian approach, and, consequently, the spectral slope corresponding to lower $k$ 
is steeper in the SMHS case, and no longer clearly consistent with the Kazantsev scaling, $k^{3/2}$. 
At the higher wavenumbers, SMHS shows more power and energy and hence a less steep spectral slope is observed than in the Cartesian approach.

A contrasting interpretation of the same CR is reported by \cite{Pipin2019}. They find this CR to be a 
violation of HSR since its small-scale magnetic helicity density in the southern hemisphere is negative.
They attribute this to the emergence and long-term presence of a prominent active region (NOAA 12192) with negative helicity
in the southern hemisphere. However, they separate small and large scales purely based on  azimuthal averaging. 
As a result, their definition of small-scale magnetic helicity density also includes large-scale non-axisymmetric contributions. 
We regard this definition as a possible source for the dissimilar interpretations: 
In the case of SMHS, the presence of a HSR-violating AR in this particular CR is seen
as increased power at intermediate to large scales, with the sign agreeing with the low-$k$
(large scale) part of the helicity spectrum. This highlights that categorising a CR as a violation of HSR or not is a delicate issue. 
SMHS offers a much richer picture, and hence can be regarded as a better-suited tool
for such classification.

\section{Discussion} \label{discuss}

Our aim with this study was to extend the Cartesian formalism for inferring spectral distributions 
of magnetic energy and helicity using the two-point correlation tensor to spherical geometry. 
The need for this was alluded to in \cite{Bran_adv}, since the analysis of \cite{Brantwo} maps the 
magnetic field vector data from solar observations to a 2D Cartesian surface. Apart from loosening
this restriction, going to spherical geometry additionally allows us to 
infer helicity spectra directly from full disk magnetograms, instead of waiting for the build-up of a synoptic map for one Carrington rotation.
Analogously, using ZDI based magnetic field 
observations \cite[for example][]{Vidotto} 
such spectra can also be retrieved for other stars limiting 
to lower spherical harmonic degree, similar to the study of \cite{Lund} who reported average magnetic 
helicity density using stellar ZDI data. It also enables
comparison with state-of-the-art dynamo models of the Sun or stars, which mostly solve the induction
equation for $\BBB$. Using our formalism one can readily compute magnetic helicity spectra without 
explicitly computing $\AAA$. 
For most of the above mentioned observations and models, a change in sign of kinetic and magnetic helicity across 
the equator owing to the Coriolis force breaking reflectional symmetry is expected.
Our formalism in Sect.~\ref{theory_sph}(b) \cite[also suggested by][]{Bran_global} by correlating fields at 
spherical harmonic degrees shifted by one, is particularly suited for such cases. Our tests in Sects.~\ref{test_dyn} 
and \ref{test_obs} confirm its applicability to inhomogeneous systems. Thus we can 
confirm and extend findings of previous studies focusing on the HSR, reported, e.g., in \cite{Singh,Pipin2019}.

Cross-helicity is also expected to play a role in generating large-scale magnetic fields \citep{Yokoi}. Previous studies 
have determined cross helicity from observations \cite[][and references therein]{Zhang_2018} mostly 
with the line-of-sight component of the velocity field. However, information about the full velocity 
field vector can also be obtained for a significant portion of the full disk as demonstrated in \cite{Rincon}.
This velocity field vector can be used with our method to infer cross-helicity spectra ($\uuu\cdot\bbb$).

\section{Conclusions}\label{conclude}
In order to investigate mechanisms responsible for large-scale magnetic fields present in the Sun
and other stars, having a knowledge of magnetic helicity and specifically its 
distribution over scales is of importance. In this study, we demonstrate an extension of an existing Cartesian formalism, 
relying on the two-point correlation tensor of the magnetic field, to spherical geometry. This allows us access 
to magnetic helicity spectra in a gauge-invariant manner by appealing to a more fundamental definition 
of helicity in terms of the linkage of the magnetic field lines. We tested this approach on a 
variety of illustrative examples before demonstrating its application to the solar vector synoptic maps.
This enables an extensive analysis of different datasets from different instruments, to vet the robustness 
of the bihelical nature of solar magnetic field against instrumental effects.

Our approach naturally captures the bihelicity of the magnetic field in each hemisphere, and also the slow modulation of the helicity as function of latitude. 
This is relevant for the Sun, where magnetic helicity at both, large and small scales, is expected to change sign across the equator.
The true nature of the helicity distribution over the whole sphere is revealed, and a potential contamination of  power and sign of the helicity from  both hemispheres is avoided. 
Such a contamination could easily lead to an apparent violation of the HSR even when the rule is obeyed.
The method discussed here remedies this, and it is expected to find applications in systems involving such rich distribution of magnetic helicity.

\begin{acknowledgements} 
MJK acknowledges the support of 
the Academy of Finland ReSoLVE Centre of Excellence (grant No.~307411).
AP was funded by the International Max Planck Research School 
for Solar System Science at the University of G\"ottingen.
This project has received funding from the European Research Council
under the European Union's Horizon 2020 research and innovation 
programme (project "UniSDyn", grant agreement n:o 818665).
SOLIS data used here are produced cooperatively by NSF/NSO and NASA/LWS.
\end{acknowledgements}

   \bibliographystyle{aa} 
   \bibliography{lit}

\begin{appendix}

\section{Spherical Correlation Function}
\label{app1}

Here we show a brief derivation of how the magnetic energy and 
magnetic helicity spectra can be extracted from the two-point
angular correlation function. Note that we are interested to determine
these spectra from the measurements of vector magnetic field from the
surface of a sphere, e.g., the Sun.
We may write the following expression for the two-point angular
correlation function under the homogeneous conditions \citep{Peebles}:
\EQ
M_{ij}(\chi)= \frac{\iint B_i(\OO_1) B_j(\OO_2)
\delta(\cos\chi_{12}-\cos\chi) ~\dd\OO_1 \dd\OO_2}
{\iint \delta(\cos\chi_{12}-\cos\chi) ~\dd\OO_1 \dd\OO_2}
\label{hom_cf1}\,,
\EN
where $\chi_{12}$ is the angle between the directions
$\OO_1=\OO_1(\theta_1,\phi_1)$ and $\OO_2=\OO_2(\theta_2,\phi_2)$
which are the position vectors of the two points on the spherical
surface; $\theta$'s and $\phi$'s represent the colatitude and azimuth,
respectively. The homogeneity is ensured by the delta function in
\Eq{hom_cf1} where the correlation function depends only on
the angular separation $\chi$, and the angular integrals yield an
average over the sphere. The normalization in the denominator may
be determined by expanding the delta function in \Eq{hom_cf1}
in terms of the Legendre Polynomials as
\EQ
\delta(\cos\chi_{12}-\cos\chi) = \sum_\ell a_\ell P_\ell(\cos\chi_{12})\,.
\EN
Writing the orthonormality condition for $P_\ell$ as
\EQ
\int_{-1}^1 P_\ell(\cos\alpha) P_{\ell'}(\cos\alpha) \dd \cos\alpha =
\frac{2\delta_{\ell \ell'}}{2\ell+1}\,,
\label{Porth}
\EN
the coefficients $a_\ell$ are found to be
\EQ
a_\ell = \frac{2\ell + 1}{2} P_\ell(\cos\chi)\,,
\EN
giving
\EQ
\delta(\cos\chi_{12}-\cos\chi) = \sum_\ell \frac{2\ell + 1}{2}
P_\ell(\cos\chi) P_\ell(\cos\chi_{12})\,.
\label{del_fn}
\EN
By expanding $P_\ell(\cos\chi_{12})$ in terms of the spherical harmonics as
\EQ
P_\ell (\cos{\chi_{12}}) = \frac{4\pi}{2\ell+1}\sum_{m=-\ell}^\ell
\ylm(\OO_1) \cylm(\OO_2)\,,
\label{Pchi12}
\EN
we find, after straightforward algebra, that
\EQ
\iint \delta(\cos\chi_{12}-\cos\chi) ~\dd\OO_1 \dd\OO_2 = 8\pi^2,
\label{norm}
\EN
for which, we have used $\sqrt{4\pi}\, Y_0^0 = 1$,
$\int\ylm(\OO) \dd \OO = \sqrt{4\pi}\delta_{\ell 0}\delta_{m 0}$ and
$P_0(\cos\chi)=1$. This gives \Eq{corr_fn}.

\section{Magnetic Energy Spectrum}
\label{app2}

Analogous to the Fourier case, we can define the magnetic
energy spectrum, which gives a distribution of the magnetic energy
over $\ell$, as
\EQ
2E_{\rm M}(\ell) = 2\pi (2\ell+1) \int_{-1}^1 \delta_{ij} M_{ij} (\chi)
P_{\ell}(\cos\chi) \dd\cos\chi\,,
\label{en_sp1}
\EN
which essentially depends on the trace of the two-point function,
${\rm Tr}[M_{ij}(\chi)]=\delta_{ij} M_{ij}$.
\Equ{en_sp1} may be understood by first expanding this trace as
$\delta_{ij} M_{ij}(\chi) = \sum_\ell c_\ell P_\ell(\cos\chi)$,
and then determining the coefficient $c_\ell$ in a standard way, which
gives
\EQ
c_\ell = \frac{2\ell+1}{2}\int_{-1}^1 \delta_{ij} M_{ij} (\chi)
P_{\ell}(\cos\chi) \dd\cos\chi
\EN
By defining the magnetic energy spectrum, $E_{\rm M}(\ell)$,
in terms of $c_\ell$ as $2E_{\rm M}(\ell) = 4\pi c_\ell$, we
arrive at \Eq{en_sp1} which, after using \Eq{corr_fn} together with
Eqs~(\ref{del_fn}), (\ref{Pchi12}) and (\ref{Porth}),
gives the following simple expression for
the energy spectrum:
\EQ
2E_{\rm M}(\ell) = \sum_{m=-\ell}^\ell (b_r^{\ell m})^2 +
(b_\theta^{\ell m})^2 + (b_\phi^{\ell m})^2\,,
\label{en_sp2}
\EN
where $b_i^{\ell m} = \int B_i(\OO) \cylm(\OO) \dd \OO$ is the expansion
coefficient of the $i^{\rm th}$ component of the magnetic field $\BBB$.

\section{Magnetic Helicity Spectrum}
\label{app3}

Gauss's linking formula yields the magnetic helicity directly in terms
of the magnetic field $\BBB(\xxx)$ by determining the flux-linkages:

\EQ
\mathcal{H}_\mathrm{M} = \frac{1}{4\pi} \iint
\frac{\xxx-\yyy}{|\xxx-\yyy|^3}	\cdot
	\left[\BBB(\xxx) \times \BBB(\yyy) \right] \dd^3 x \dd^3 y\,.
\EN
Here $\xxx$ and $\yyy$ are the position vectors of two points on the surface
of a sphere. Using
\EQ
\frac{\xxx-\yyy}{|\xxx-\yyy|^3}=
\bnabla_y\frac{1}{|\xxx-\yyy|}\,,\quad \mbox{and} \quad
\frac{1}{|\xxx-\yyy|}=
\frac{1}{R}\sum_{\ell =0}^\infty
P_\ell (\cos{\chi})\,,
\EN
with $R$, $\chi$, and $P_\ell$ being, respectively, the radius of the sphere,
the angle between $\xxx$ and $\yyy$, and the Legendre polynomial of degree
$\ell$, we rewrite the expression for the magnetic helicity,
\EQ
\mathcal{H}_\mathrm{M} = \sum_{\ell=0}^\infty H_M(\ell)
\EN
where
\begin{eqnarray}
	H_M(\ell)&=&\frac{1}{2\ell+1}\iint
	\bnabla' \left(\sum_{m=-\ell}^\ell \ylm(\OO) \cylm(\OO')
	\right) \,\cdot \nonumber \\[2ex] 
	&&\qquad\qquad \cdot
	\left[\BBB(\OO) \times \BBB(\OO') \right] \dd \OO  \dd \OO'
	\label{final_exp}
\end{eqnarray}
for which, we assumed a unit sphere ($R=1$) and expanded the Legendre
polynomials in terms of the spherical harmonics as
\EQ
P_\ell (\cos{\chi}) = \frac{4\pi}{2\ell+1}\sum_{m=-\ell}^\ell
\ylm(\theta,\phi) \cylm(\theta',\phi')\,.
\EN
Directions $\OO=\OO(\theta,\phi)$ and $\OO'=\OO'(\theta',\phi')$
correspond to the position vectors $\xxx$ and
$\yyy$, respectively, with $\dd \OO=\sin{\theta}\dd \theta \dd \phi$ and
$\dd \OO'=\sin{\theta'}\dd \theta' \dd \phi'$. Employing the Jeans relation
which expresses the wavenumber $k$ on the surface of the sphere of radius
$R$ (assumed as unity here) in terms of the spherical harmonic degree $\ell$
by $kR=\sqrt{\ell(\ell+1)}\approx \ell+1/2$, we can write
\EQ
kH_M(\ell)=(\ell+1/2)H_M(\ell)\,.
\label{khm}
\EN
Thus the \emph{scaled magnetic helicity
spectrum}, $kH_M(\ell)$, which has the same dimensions as for the magnetic
energy spectrum, $E_M(\ell)$, may be determined by using
\Eqs{final_exp}{khm}.

\emph{It is important to note that \Eq{final_exp} provides a general
expression for the magnetic helicity, and thus, in its current
form, it does not make any assumption of homogeneity.}
This same equation may be used for both, homogeneous and inhomogeneous,
cases by suitably writing the two-point function $M_{ij}$ which
naturally appears when we take an ensemble average of \Eq{final_exp}.
We take (i) $M_{ij} = M_{ij}(\chi)$ for homogeneous case in which
it depends only on the angular separation ($\chi$) between the
two points, and (ii) $M_{ij} = M_{ij}(\OO,\chi)$ for inhomogeneous
case where $M_{ij}$ depends also on the position ($\OO$) on the surface
of the sphere. In weakly inhomogeneous turbulence, which appears to be more
relevant in the solar context, $M_{ij}$ is expected to vary rapidly
with $\chi$ while showing a slow variation with position $\OO$ on
the sphere. Here we are more interested in only the latitudinal
variation which involves a sign change of magnetic helicity
across the equator, simultaneously at both, small and large,
length scales.

To numerically evaluate the integrals in \Eq{final_exp} we use the
vectorial generalisation of spherical harmonics implemented in the
SHTns library \citep{Schaeffer}.
\end{appendix}
\end{document}